\documentclass[preprint,12pt]{elsarticle}

\usepackage{graphicx} 
\usepackage{dcolumn} 
\usepackage{bm} 
\usepackage{amsmath}
\usepackage{braket}
\usepackage[normalem]{ulem}
\usepackage[hidelinks]{hyperref}
\usepackage{epstopdf}
\usepackage{placeins}
\usepackage{multirow}
\usepackage{makecell}
\usepackage{cleveref}
\usepackage{xcolor}

\newcommand{\beq}{\begin{eqnarray}}
\newcommand{\eeq}{\end{eqnarray}}
\newcommand{\beqnn}{\begin{eqnarray*}}
\newcommand{\eeqnn}{\end{eqnarray*}}

\newcommand{\Tr}{\mathrm{Tr}}
\newcommand{\SO}{\mathrm{SO}}
\newcommand{\SU}{\mathrm{SU}}

\def\spose#1{\hbox to 0pt{#1\hss}}
\def\ltapprox{\mathrel{\spose{\lower 3pt\hbox{$\mathchar"218$}}
 \raise 2.0pt\hbox{$\mathchar"13C$}}}

\journal{Physics Letters B}

\begin{document}
	
\begin{frontmatter}

\title{Towards glueball masses of large-$N$ $\mathrm{SU}(N)$ pure-gauge theories without topological freezing}

\author[a]{Claudio Bonanno\corref{cor}}
\cortext[cor]{Corresponding author}
\ead{claudio.bonanno@fi.infn.it}
\affiliation[a]{organization={INFN Sezione di Firenze},
addressline={Via G.~Sansone 1},
city={Sesto Fiorentino, Firenze},
postcode={I-50019},
state={Italia}}

\author[b]{Massimo D'Elia}
\ead{massimo.delia@unipi.it}
\affiliation[b]{organization={Universit\`a di Pisa and INFN Sezione di Pisa},
addressline={Largo Pontecorvo 3},
city={Pisa},
postcode={I-56127},
state={Italia}}

\author[c]{Biagio Lucini}
\ead{b.lucini@swansea.ac.uk}
\affiliation[c]{organization={Department of Mathematics, Faculty of Science and Engineering, Swansea University},
addressline={Fabian Way},
city={Swansea},
postcode={SA1 8EN},
state={Wales, UK}}

\author[d,e]{Davide Vadacchino}
\ead{davide.vadacchino@plymouth.ac.uk}
\affiliation[d]{organization={School of Mathematics and Hamilton Mathematics Institute, Trinity College},
state={Dublin 2, Ireland}}
\affiliation[e]{organization={Centre for Mathematical Sciences, University of Plymouth},
city={Plymouth},
postcode={PL4 8AA},
state={UK}}

\date{\today}

\begin{abstract}
In commonly used Monte Carlo algorithms for lattice gauge theories the integrated autocorrelation time of the topological charge is known to be exponentially-growing as the continuum limit is approached. This \emph{topological freezing}, whose severity increases with the size of the gauge group, can result in potentially large systematics. To provide a direct quantification of the latter, we focus on $\mathrm{SU}(6)$ Yang--Mills theory at a lattice spacing for which conventional methods associated to the decorrelation of the topological charge have an unbearable computational cost. We adopt the recently proposed \emph{parallel tempering on boundary conditions} algorithm, which has been shown to remove systematic effects related to topological freezing, and compute glueball masses with a typical accuracy of $2-5\%$. We observe no sizeable systematic effect in the mass of the first lowest-lying glueball states, with respect to calculations performed at nearly-frozen topological sector.
\end{abstract}


\end{frontmatter}

\newpage

\section{Introduction}
Based on the confining properties of QCD, it is predicted that gauge-invariant bound states made of gluons alone, called \emph{glueballs}, should appear in the spectrum as asymptotic states. So far, glueballs have eluded experimental detection, although candidate events have been recently found~\cite{TOTEM:2020zzr}. From the theoretical point of view, during last decades several studies have appeared in the literature where glueball masses are computed in the non-perturbative setting provided by numerical lattice field theory simulations, where many intriguing predictions can been derived, both in relation to QCD~\cite{Lucini:2001ej,Lucini:2004my,Lucini:2010nv,Hong:2017suj,Bennett:2020hqd,Athenodorou:2020ani,Athenodorou:2021qvs} or to possible Standard Model extensions~\cite{Bennett:2020qtj}. Regarding QCD glueballs, the majority of such predictions has been obtained for quarkless pure-gauge theories in the large number of colors ($N$) limit $N\to \infty$. Large-$N$ pure $\SU(N)$ gauge theories provide a reasonable approximation of real-world $N=3$ QCD~\cite{Lucini:2012gg}, as finite-$N$ corrections are suppressed as powers of $1/N^2$, and enable us to avoid some technical complications (e.g., all glueballs are exactly non-interacting and have infinite lifetime at $N=\infty$).\\
\indent The extraction of glueball masses from lattice simulations at large $N$ is a non-trivial task and several sources of systematic errors have to be addressed to obtain reliable results. In the last decades, enormous progress has been made in the development and the refinement of the relevant techniques~\cite{BERG1983109, APE:1987ehd, TEPER1987345, Morningstar:1997ff, Teper:1998te,Morningstar:1999rf, Lucini:2001ej, Lucini:2004my, Blossier:2009kd,Lucini:2010nv}. Nevertheless, systematic effects related to \emph{topological freezing}~\cite{Schaefer:2010hu} have never been addressed in a satisfactory way, so far.\\
\indent Standard local updating algorithms suffer from non-ergodicity in the 
vicinity of the continuum limit: as $a\to0$, the  Markov chain of configurations
explored by the system tends to remain trapped in a fixed topological sector. This problem becomes exponentially more severe as $N$ is increased. When $N$ is large, the evolution of the topological charge along the Monte Carlo trajectory freezes already for coarse lattice spacings~\cite{Lucini:2004yh,DelDebbio:2004xh, DelDebbio:2006yuf, Bonati:2016tvi}. In particular, there is ample numerical evidence that the autocorrelation time of the topological charge $\tau(Q)$ diverges exponentially as a function of $1/a$ and/or $N$~\cite{DelDebbio:2004xh, DelDebbio:2006yuf, Bonati:2016tvi}, thus making ergodic exploration of different topological sectors rapidly unfeasible as the continuum limit and/or the large-$N$ limit are approached. In practice, when $N \ge 6$ and $a \sim 0.1$~fm or below, essentially very few to no fluctuations of $Q$ are observed during reasonably-long Monte Carlo histories\footnote{An exploratory study of the topological charge in the theory with dynamical fermions for $3 \le N \le 5$ has been performed in~\cite{DeGrand:2020utq}.}.\\
\indent Since there is theoretical evidence that computing glueball masses on a fixed topological sector may introduce a bias~\cite{Brower:2003yx}, it is of utmost importance to check that any systematic error related to the restriction in a fixed topological sector is under control within the typical precision achieved in actual simulations. The effect of fixed topology has however never been systematically probed on large-$N$ glueball mass computations from the lattice.\\
\indent In this letter we make a first step in this direction by removing any systematic effect related to the freezing problem through the \emph{Parallel Tempering on Boundary Conditions} (PTBC). The PTBC algorithm was proposed by M.~Hasenbusch~\cite{Hasenbusch:2017unr} for $2d$ large-$N$ $\mathit{CP}^{N-1}$ models and was recently employed both in the latter case and in large-$N$ $\SU(N)$ pure-gauge theories~\cite{Berni:2019bch,Bonanno:2020hht} to improve state of the art of large-$N$ topology from the lattice. In particular, the PTBC algorithm has been shown to provide a dramatic enhancement compared to standard algorithms when looking at the evolution of the topological charge $Q$, allowing to achieve a gain of several orders of magnitude in terms of $\tau(Q)$.\footnote{Other recently proposed algorithms to avoid topological freezing include~\cite{Cossu:2021bgn}.}\\
\indent We generated 20k well-decorrelated gauge configurations for $\SU(6)$ at $a \simeq 0.0938$~fm adopting the PTBC algorithm. These configurations were then used to compute glueball masses for the first few lightest states using standard methods. Since the PTBC algorithm is designed to restore ergodicity, our simulations frequently explored $Q\ne0$ sectors. Hence, for our model, we were able to provide the first results for glueball masses free of any systematics related to topological freezing.\\
\indent This letter is organized as follows: in Sec.~\ref{sec:setup} we describe our numerical setup, in Sec.~\ref{sec:results} we show our results for glueball masses and compare them with results obtained with standard algorithms, finally in Sec.~\ref{sec:conclusions} we draw our conclusions.

\section{Lattice setup}\label{sec:setup}

We consider a collection of $N_r$ hypercubic lattice replicas with $L^4$ sites. Replicas differ from one another only in the boundary conditions imposed on the links on a small sub-region of the lattice, $D$, which we call \emph{the defect}. Boundary conditions on the defect are chosen in order to interpolate between open boundary conditions (OBC)~\cite{Luscher:2011kk} and periodic boundary conditions (PBC), while are taken periodic elsewhere for every replica. Each replica is evolved independently using standard local algorithms. After each replica has been updated, swaps among different replicas are proposed and accepted/rejected by means of a standard Metropolis test. Iterations over the full lattice are alternated with hierarchical updates over small sub-lattices centered around $D$ to improve the efficiency of the algorithm.

In practice, the lattice action of the $r^{\text{th}}$ replica looks like
\begin{align*}
S_L^{(r)} = -\frac{\beta}{N} \sum_{x, \mu>\nu} K^{(r)}_{x,\mu} K^{(r)}_{x+\hat{\mu},\nu} K^{(r)}_{x+\hat{\nu},\mu} K^{(r)}_{x,\nu} \,\, \Re \Tr \Pi^{(r)}_{x,\mu\nu},
\end{align*}
where $\beta$ is the bare coupling, $\Pi^{(r)}_{x,\mu\nu}$ is the plaquette computed on the gauge configuration of the $r^{\text{th}}$ replica and 
\beqnn
K_{x,\mu}^{(r)} =
\begin{cases}
c(r), \quad &\mbox{ if} \quad \mu = 1 \ \mathrm{and}\ x \in D, \\
1, \quad &\mbox{ otherwise,}
\end{cases}
\eeqnn
is used to impose boundary conditions on the links crossing orthogonally the defect $D=\{x_1=L(a-1), \, 0\le x_2 < L_d^{(2)}, \, 0\le x_3 < L_d^{(3)}, \, 0\le t < L_d^{(4)}\}$. In our simulations we used $L_d^{(2)} = L_d^{(3)} = L_d^{(4)} \equiv L_d$, and the defect is kept fixed in the position described here; its position is however effectively moved by translating the periodic copy (which is translation-invariant). Coefficients $c(r)$ interpolate between $c(0)=1$ (PBC) and $c(N_r-1)=0$ (OBC) and are tuned through short runs to make swap probabilities uniform among different replicas.

Glueball masses are computed on the periodic $r=0$ replica using standard techniques, which we here succinctly summarize. We define a variational basis $\mathcal{B} = \{O_i(t)\}$ of time-dependent operators, with quantum numbers compatible with the desired glueball state. We only consider zero-momentum operators $O_i(t) = \sum_{\vec{x}} O_i(t,\vec{x})$, where $O_i(t,\vec{x})$ are gauge-invariant 
local operators expressed as traces of products of links 
taken over closed space-like lattice paths. 
We also include in $\mathcal{B}$ operators obtained from blocked and smeared links. Once $\mathcal{B}$ is chosen, to extract the lightest state in the selected channel, we compute $C_{ij}(t) = \braket{O_i(t) O_j(0)}$ and, through the \emph{Generalized EigenValue method} (GEV), we obtain the eigenvector $\overline{v}_i$ related to the largest eigenvalue of the generalized eigenvalue problem $C_{ij}(t) v_j = \lambda(t,t^\prime)C_{ij}(t^\prime)v_j$. The correlator of the best overlapping operator between the vacuum and the desired glueball state is then obtained as $ C_{\text{best}}(t) \equiv C_{ij}(t) \overline{v}_i \overline{v}_j$. The glueball mass $m$ is finally obtained in lattice units through a best fit of the expression
\beq\label{eq:correlator_mass_fit}
C_{\text{best}}(t) \sim \exp(-amt),
\eeq
where the fit is performed over a range where the effective mass
\beq\label{eq:effective_mass}
am_{\mathrm{eff}}(t) \equiv - \log \left( \frac{ C_{\text{best}}(t+a) }{ C_{\text{best}}(t) } \right)
\eeq
shows a plateau. For more details about the glueball mass extraction procedure, the choice of the variational basis for each $\mathcal{B}$ channel and the smearing algorithms adopted in the context of glueball mass computations, we refer, e.g., to Refs.~\cite{BERG1983109,APE:1987ehd,TEPER1987345,Teper:1998te,Lucini:2001ej,Lucini:2004my,Blossier:2009kd,Lucini:2010nv,Bennett:2020qtj,Athenodorou:2020ani,Athenodorou:2021qvs}.

\section{Results}\label{sec:results}
We simulated $\SU(6)$ at $\beta=25.452$ ($a\simeq 0.0938$~fm) on a $16^4$ lattice with a cubic $L_d=3$ defect. For each of $10$ independent runs, we collected $2000$ well-decorrelated configurations, stored every $200$ parallel tempering steps, after discarding the first $10000$ parallel tempering steps for thermalization. A single parallel tempering step is performed as follows:
\begin{enumerate}
\item Each replica is updated in parallel with a full lattice sweep of a $4$:$1$ combination of over-relaxation and over-heat-bath algorithms (in the following, this combination will be referred to as ``standard updating step'').
\item\label{enum:swaps} Swaps are proposed between replica pairs $(r,r+1)$, first for $r$ even, then for $r$ odd or viceversa (order decided stochastically). Swaps of odd and even $(r,r+1)$ pairs are proposed in parallel and accepted with probability
\begin{align*}
p(r,r+1) = \min\left\{1, \exp(-S_L(r\leftrightarrow r+1) + S_L(\text{no swap})) \right\}
\end{align*}
After the swap proposals, the periodic $r=0$ replica is translated of $1$ lattice site along a random direction to effectively move the position of the defect.
\item Each replica is updated in parallel with hierarchical sweeps on small sub-lattices centered around the defect. After each hierarchical iteration, the swaps and the $r=0$ replica translations are performed as in~\ref{enum:swaps}.
\end{enumerate}
It is clear that a single parallel tempering step requires a factor of $\sim N_r$ larger numerical effort compared to a standard updating step. Nonetheless, even when considering this overhead, the obtained gains in terms of decorrelation of the topological charge make the parallel tempering algorithm the obvious method of choice between the two, as it will be manifest in the following. With this implementation, collecting the sample of configurations employed 
for this study required $\sim 2.3$M core-hours on \texttt{Intel Skylake} 
processors. 

\begin{figure}[!htb]
\centering
\includegraphics[scale=0.38]{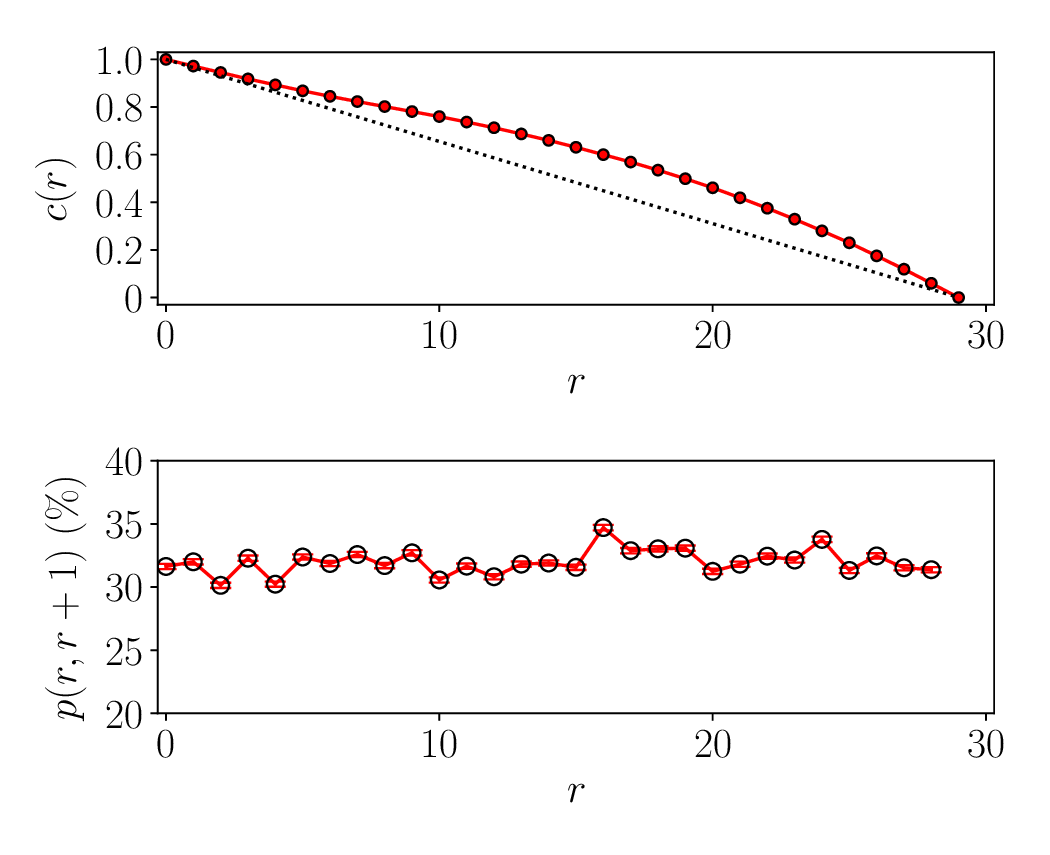}
\includegraphics[scale=0.38]{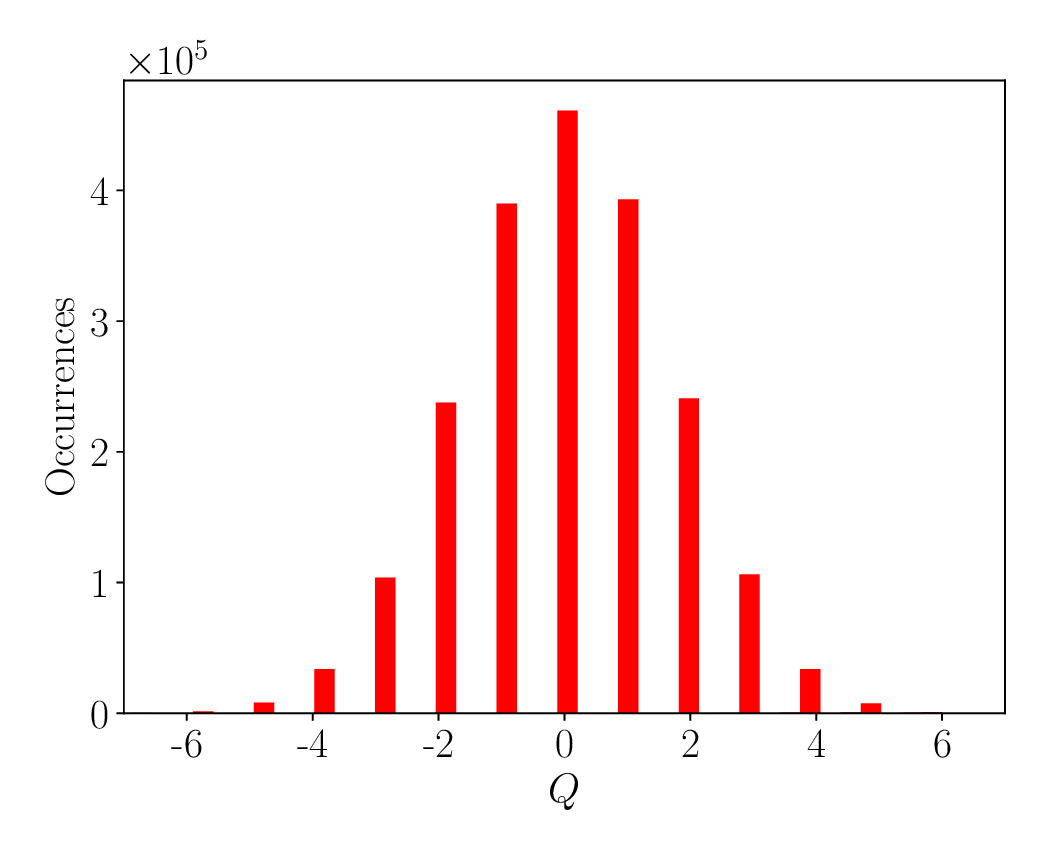}
\includegraphics[scale=0.38]{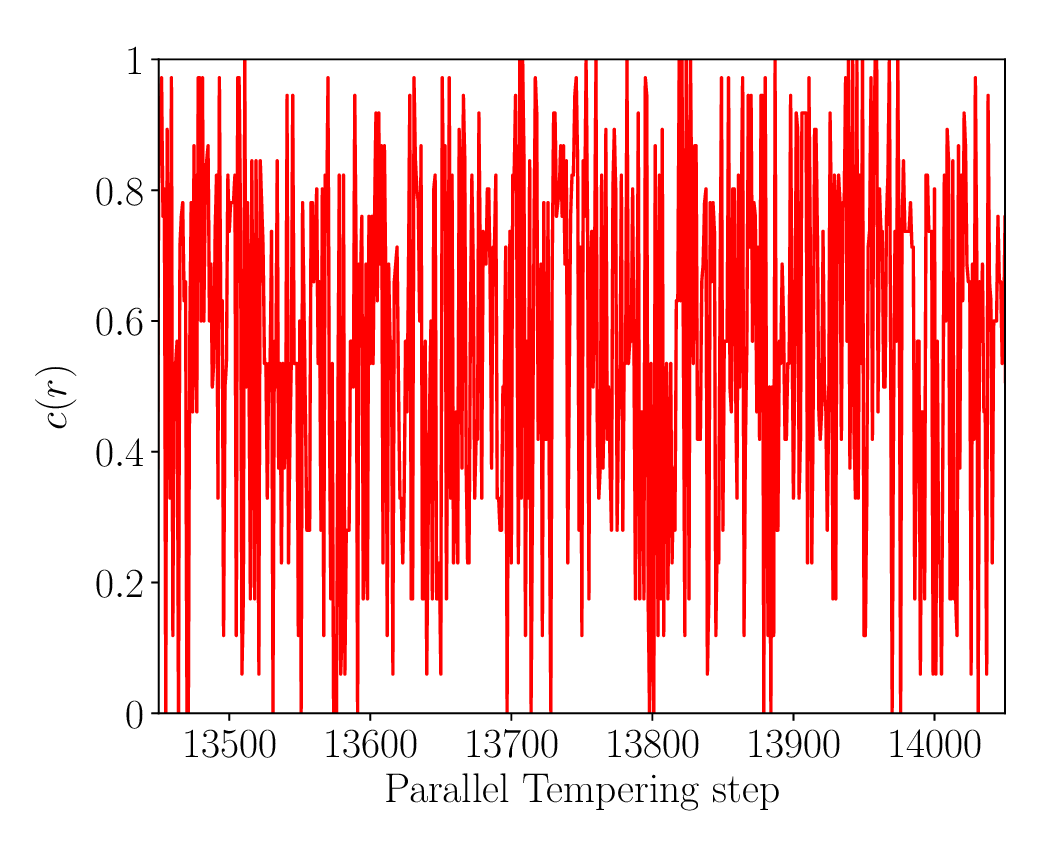}
\includegraphics[scale=0.38]{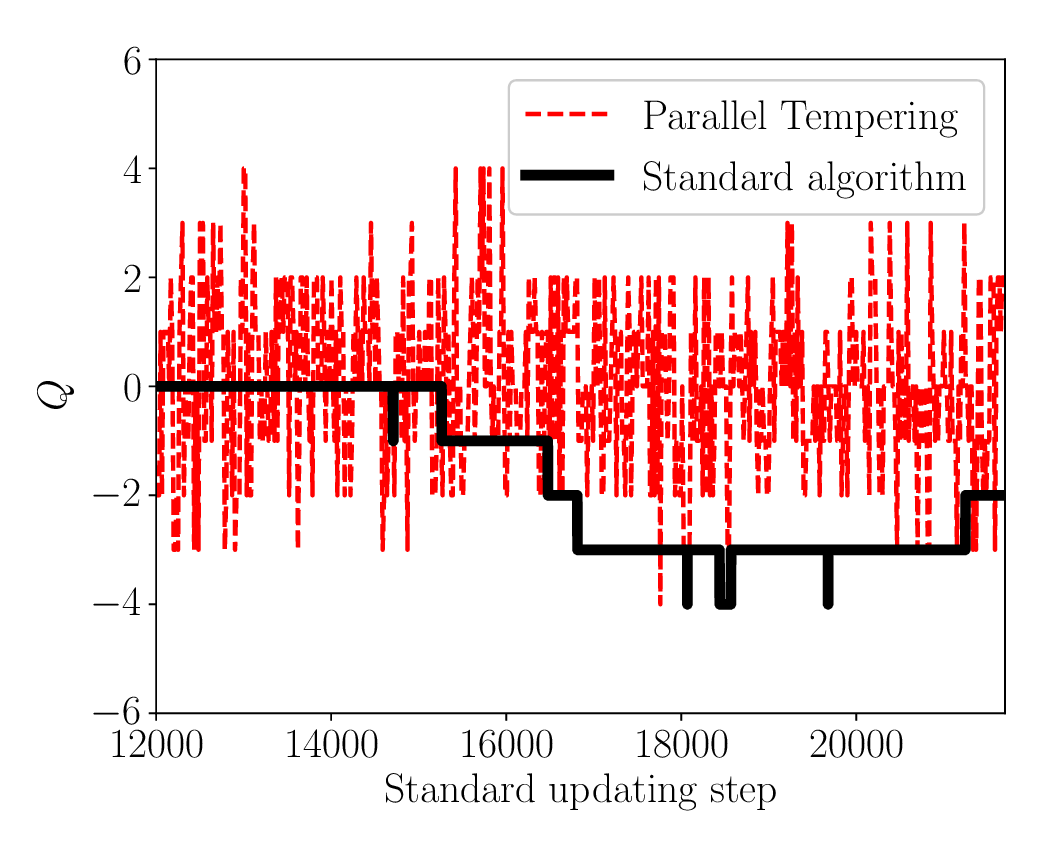}
\caption{Left plot above: choice of $c(r)$ and related swap probabilities. Dotted line represents naive uniform choice of $c(r)$. Right plot above: histogram of the topological charge obtained from our configuration sample, generated with the PTBC algorithm. Left plot below: random walk of a configuration through different replicas. Time along horizontal axis is expressed in units of PTBC steps, and the shown time window corresponds to $\sim10^{-2}\%$ of our total statistics. Right plot below: Monte Carlo evolution of the topological charge $Q$ obtained with the PTBC and with the standard algorithms. The time on the horizontal axis is expressed in units of standard updating steps for both algorithms (Monte Carlo time of the PTBC run was rescaled with a factor of $N_r$), and the shown time window corresponds to $\sim 0.2\%$ of the total statistics collected with the PTBC algorithm.}
\label{fig:details_parallel_tempering}
\end{figure}

We chose a uniform $\sim30\%$ swap probability for all pairs; to reach it we needed $N_r = 30$ replicas. In Fig.~\ref{fig:details_parallel_tempering} we show the behavior of $c(r)$ for every $r$ and the related swap acceptances (left plot above). Choosing approximately uniform swap probabilities among different replica pairs ensured that a given configuration explored uniformly all boundary conditions $c(r)$ in a random-walk fashion, which is a necessary condition for the correct operation of the PTBC algorithm (left plot below).

Moreover, In Fig.~\ref{fig:details_parallel_tempering} we also show the histogram of the obtained sampling of the topological charge $Q$ (right plot above) and the history of the topological charge evolution in our typical run compared to the evolution obtained with standard algorithms (right plot below). This quantity was computed from the standard clover definition on smoothened configurations, obtained after $20$ cooling steps, and rounded to the nearest integer using the so-called \emph{alpha-rounding} method explained in, e.g., Refs.~\cite{DelDebbio:2002xa,Bonati:2016tvi,Bonanno:2020hht}. The PTBC algorithm is capable of performing an ergodic sampling of the space of configurations with respect to the topological charge, and allows to observe numerous fluctuations of $Q$ in a case where, with standard algorithms, only a handful would be observed, cfr.~Fig.~\ref{fig:details_parallel_tempering}. The gain in terms of the integrated autocorrelation time of the topological charge $\tau(Q)$ is dramatic: while for the standard run we estimate $\tau_{\mathrm{std}}(Q) \sim 5000$, with the PTBC algorithm we find $\tau_{\mathrm{PTBC}}(Q) = 92(8)$ (where $\tau_{\mathrm{PTBC}}$ was obtained keeping into account that a single PTBC step requires a numerical effort which is larger by approximately a factor of $\sim N_r$ compared to a standard updating step).

We then employed the generated sample to compute glueball masses for the Ground State (GS) of all $\mathrm{R}^{\mathrm{PC}}$ channels, with the exception of $\mathrm{A}_1^{--}$ and $\mathrm{A}_2^{-+}$, which appear to be heavier than our ultra-violet cut-off $\Lambda_{\mathrm{UV}} \sim 2/a$, and defined the dimensionless ratios $m_{\mathrm{R}^{\mathrm{PC}}}/m_{\mathrm{A}_1^{++}} \sim m_{J^{\mathrm{PC}}}/m_{0^{++}}$. Here $\mathrm{R}$ stands for a particular representation of the octahedral group, $J$ stands for the corresponding representation of $\SO(3)$ in the continuum, and $\mathrm{PC}$ stands for the \emph{spatial parity} and \emph{charge conjugation} quantum numbers. For the ground states in the $\mathrm{R}^{\mathrm{PC}}$ channels, we can establish the following correspondence among representations of the octahedral group and representations of $\SO(3)$: $\mathrm{A}_1 \rightarrow J=0$, $\mathrm{A}_2 \rightarrow J=3$, $\mathrm{E} \rightarrow J=2$, $\mathrm{T}_1 \rightarrow J=1$, $\mathrm{T}_2 \rightarrow J=2$.

The ratios $m_{\mathrm{R}^{\mathrm{PC}}}/m_{\mathrm{A}_1^{++}}$ were then obtained with a precision of the order of $2-5\%$. 
We then compared our results for such quantities with those reported in Ref.~\cite{Athenodorou:2021qvs}, using $a m_{\mathrm{A}_1^{++}}$ to fix a common lattice spacing scale. As the sample of configurations analyzed in Ref.~\cite{Athenodorou:2021qvs} was obtained from standard local algorithms, it is affected by severe topological freezing. 
The comparison was performed as follows. We first extrapolated the finite-$a$ results of Ref.~\cite{Athenodorou:2021qvs} towards the continuum limit by fitting
\beq\label{eq:continuum_limit}
\frac{m_{\mathrm{R}^{\mathrm{PC}}}}{m_{\mathrm{A}_1^{++}}}(a) = \frac{m_{J^{\mathrm{PC}}}}{m_{0^{++}}} + c_{\mathrm{R}^{\mathrm{PC}}} \left(am_{\mathrm{A}_1^{++}}\right)^2 + o(a^2);
\eeq
to the data, using $c_{\mathrm{R}^{\mathrm{PC}}}$ and $m_{J^{PC}}/m_{0^{++}}$ as fitting parameters. Then, we computed the value of $m_{\mathrm{R}^{\mathrm{PC}}}/m_{\mathrm{A}_1^{++}}$ expected at our value of $am_{\mathrm{A}_1^{++}}$, according to the best fit of Eq.~\eqref{eq:continuum_limit} above. To allow a comparison, we report in Tab.~\ref{tab:hpt_results} our results for the mass of the GS in each $\mathrm{R}^{\mathrm{PC}}$ channel, expressed in lattice units, as well as our determinations for $m_{\mathrm{R}^{\mathrm{PC}}}/m_{\mathrm{A}_1^{++}}$. The determinations from Ref.~\cite{Athenodorou:2021qvs}, using the procedure described above, are also reported in Tab.~\ref{tab:hpt_results}, in the rightmost column.

\begin{table}[!htb]
\begin{center}
\begin{tabular}{|c|c||c|c|c|}
\hline
& & & &\\[-1em]
$\mathrm{R}^{\mathrm{PC}}$ & $J^{\mathrm{PC}}$& \makecell{$am_{\mathrm{R}^{\mathrm{PC}}}$ \\ \\ this work \\ $\beta=25.452$} & \makecell{$m_{\mathrm{R}^{\mathrm{PC}}}/m_{\mathrm{A}_1^{++}}$ \\ \\ this work \\ $\beta=25.452$} & \makecell{$m_{\mathrm{R}^{\mathrm{PC}}}/m_{\mathrm{A}_1^{++}}$ \\ \\ Ref.~\cite{Athenodorou:2021qvs} \\ interpolation in $am_{\mathrm{A}_1^{++}}=0.6246$} \\
& & & &\\[-1em]
\hline
$\mathrm{A}_1^{++}$ & $0^{++}$ & 0.6246(78) & -         & -          \\
\hline
\multicolumn{5}{|c|}{ }\\[-1em]
\multicolumn{5}{|c|}{Shown in Fig.~\ref{fig:plot_results}}\\
\hline
$\mathrm{E}^{++}$   & $2^{++}$ & 0.956(44)  & 1.530(73) & 1.5545(61) \\
$\mathrm{T}_2^{++}$ & $2^{++}$ & 0.991(16)  & 1.586(33) & 1.5720(62) \\
$\mathrm{A}_1^{-+}$ & $0^{-+}$ & 1.034(20)  & 1.655(37) & 1.6370(95) \\
$\mathrm{T}_1^{+-}$ & $1^{+-}$ & 1.194(90)  & 1.91(15)  & 1.9527(97) \\
$\mathrm{E}^{-+}$   & $2^{-+}$ & 1.281(26)  & 2.050(49) & 2.037(12)  \\
\hline
\multicolumn{5}{|c|}{ }\\[-1em]
\multicolumn{5}{|c|}{Not shown in Fig.~\ref{fig:plot_results}}\\
\hline
$\mathrm{T}_2^{-+}$ & $2^{-+}$ & 1.40(10)   & 2.23(17)  & 2.0483(92) \\
$\mathrm{A}_2^{+-}$ & $3^{+-}$ & 1.543(45)  & 2.471(78) & 2.369(19)  \\
$\mathrm{A}_2^{++}$ & $3^{++}$ & 1.548(40)  & 2.478(71) & 2.424(25)  \\
$\mathrm{T}_2^{+-}$ & $2^{+-}$ & 1.571(46)  & 2.516(80) & 2.363(14)  \\
$\mathrm{T}_2^{--}$ & $2^{--}$ & 1.578(58)  & 2.526(98) & 2.567(31)  \\
$\mathrm{T}_1^{++}$ & $1^{++}$ & 1.675(59)  & 2.68(10)  & 2.517(18)  \\
$\mathrm{E}^{--}$   & $2^{--}$ & 1.696(61)  & 2.71(10)  & 2.563(17)  \\
$\mathrm{T}_1^{--}$ & $1^{--}$ & 1.700(60)  & 2.72(10)  & 2.481(30)  \\
$\mathrm{T}_1^{-+}$ & $1^{-+}$ & 1.709(65)  & 2.74(11)  & 2.857(19)  \\
$\mathrm{E}^{+-}$   & $2^{+-}$ & 1.845(93)  & 2.95(15)  & 2.879(27)  \\
$\mathrm{A}_2^{--}$ & $3^{--}$ & 1.849(89)  & 2.96(15)  & 2.869(35)  \\
$\mathrm{A}_1^{+-}$ & $0^{+-}$ & 2.00(12)   & 3.20(20)  & 3.112(49)  \\
\hline
\end{tabular}
\end{center}
\caption{Summary of the obtained results for the mass $m_{\mathrm{R}^{\mathrm{PC}}}$ of the GS of all accessible $\mathrm{R}^{\mathrm{PC}}$ channels in lattice units for $N=6$ and $\beta=25.452$ on a $16^4$ lattice, obtained from gauge configurations generated with the PTBC algorithm. We also compare our results for the ratios $m_{\mathrm{R}^{\mathrm{PC}}}/m_{\mathrm{A}_1^{++}}$ with those obtained interpolating the best fit of Eq.~\eqref{eq:continuum_limit} to results of Ref.~\cite{Athenodorou:2021qvs} for our value of $am_{\mathrm{A}_1^{++}}$.}
\label{tab:hpt_results}
\end{table}

\begin{figure}[!htb]
\centering
\includegraphics[scale=0.38]{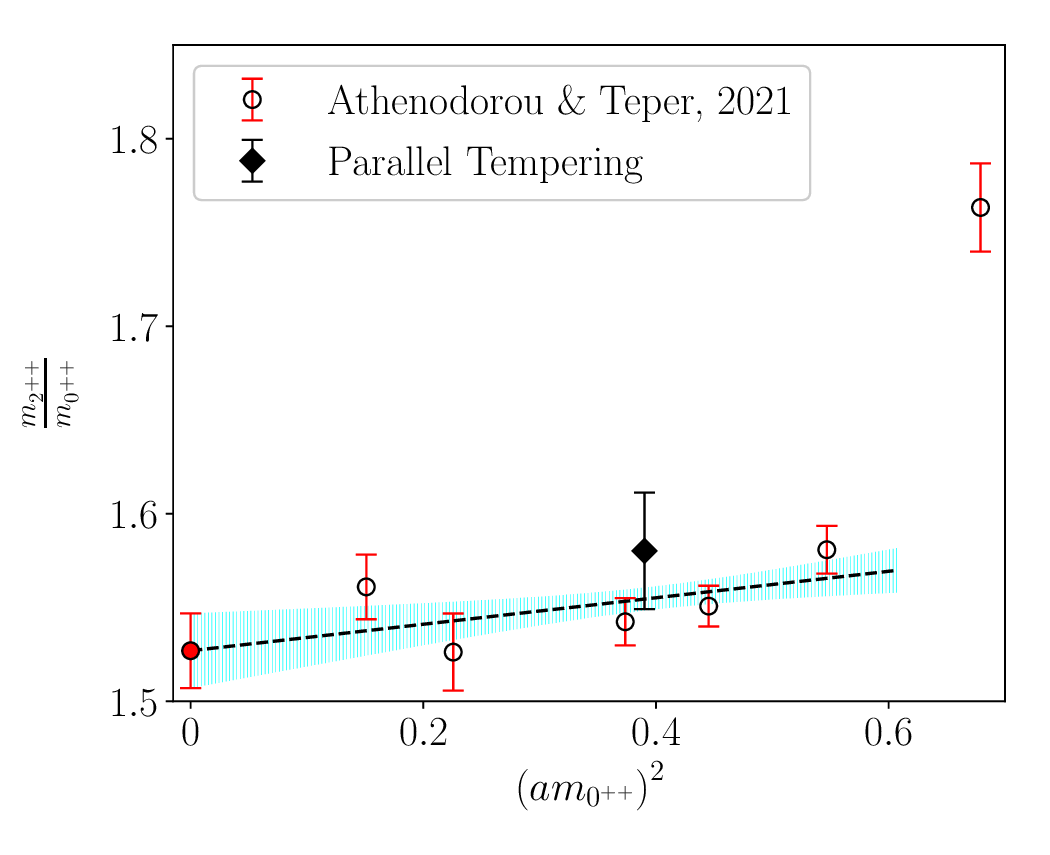}
\includegraphics[scale=0.38]{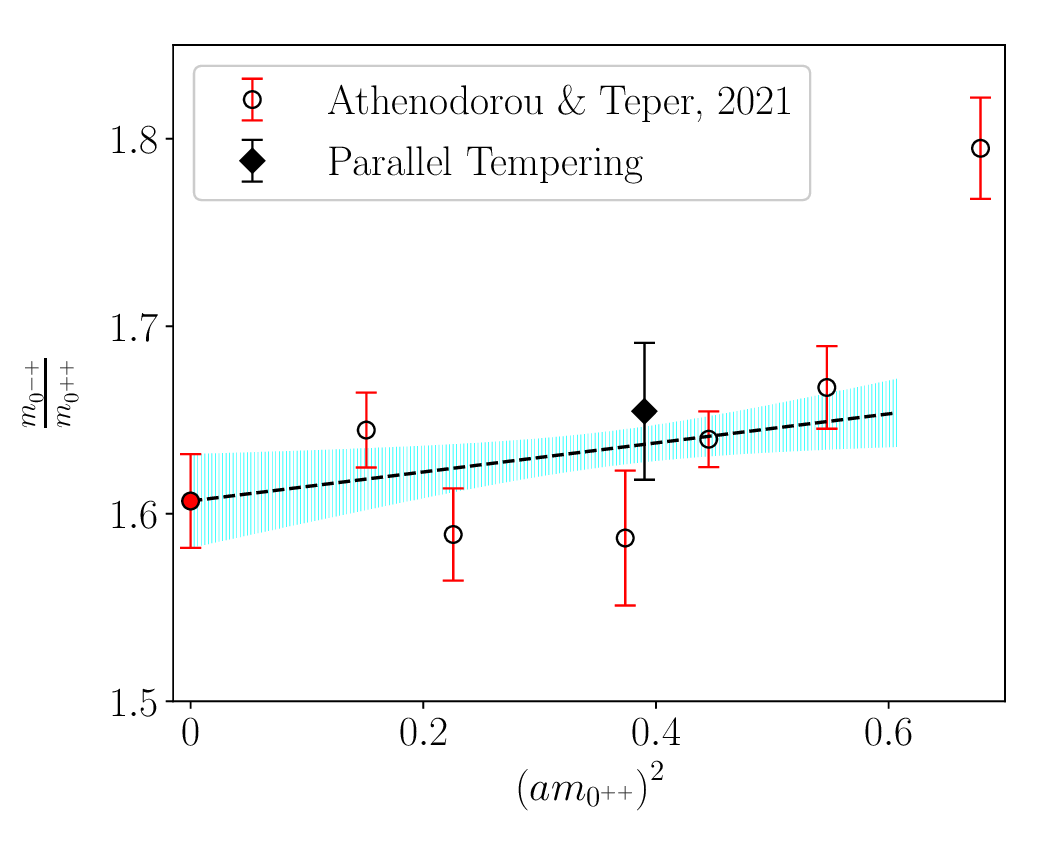}
\includegraphics[scale=0.38]{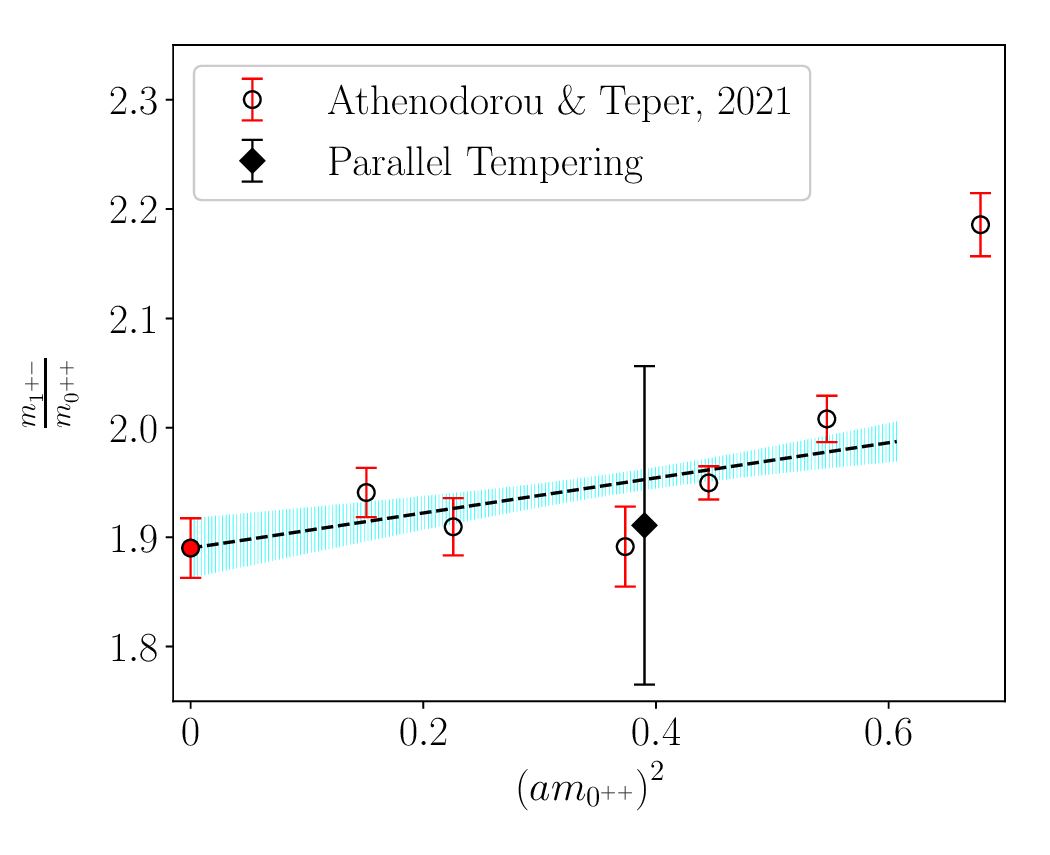}
\includegraphics[scale=0.38]{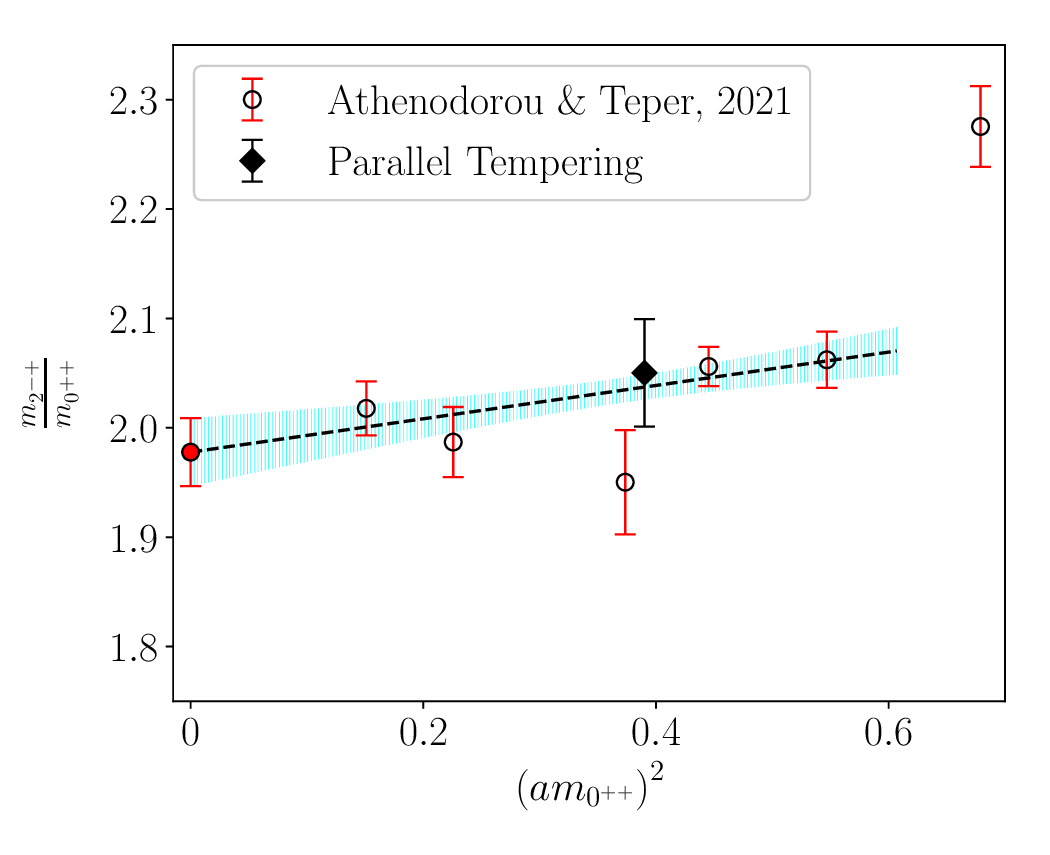}
\caption{Results for the ratios of $m_{2^{++}}$ (left-above plot), $m_{0^{-+}}$ (right-above plot), $m_{1^{+-}}$ (left-below plot) and $m_{2^{-+}}$ (right-below plot) to $m_{0^{++}}$ obtained from configurations generated with the PTBC algorithm (diamond points), compared to those of Ref.~\cite{Athenodorou:2021qvs}, obtained from configurations generated with standard local algorithms (empty round points). The correspondence among $\mathrm{R}^{\mathrm{PC}}$ and $J^{\mathrm{PC}}$ channels was done according to Tab.~\ref{tab:hpt_results}. Full round points, dashed lines and shadowed areas represent, respectively, the continuum extrapolation results, linear best fits and related fit errors of data of Ref.~\cite{Athenodorou:2021qvs}.}
\label{fig:plot_results}
\end{figure}

A comparison can also be made from Fig.~\ref{fig:plot_results}, where, for brevity, we just show the masses of the first $4$ lightest states above $0^{++}$ (the GS of $\mathrm{A}_1^{++}$): $2^{++}$ (obtained from the weighted arithmetic mean between the mass of the GS of the $\mathrm{E}^{++}$ channel and the mass of the GS of the $\mathrm{T}_2^{++}$ channel, which are expected to become degenerate in the continuum limit), $0^{-+}$ (the GS of the $\mathrm{A}_1^{-+}$ channel), $1^{+-}$ (the GS of the $\mathrm{T}_1^{+-}$ channel) and $2^{-+}$ (the GS of the $\mathrm{E}^{-+}$ channel). We stress that no difference was observed in heavier channels compared to the results we are displaying in Fig.~\ref{fig:plot_results}, cfr.~Tab.~\ref{tab:hpt_results}.

As a matter of fact, in none of the explored cases any systematic effect related to topological freezing was observed. Our results always fall on top of the ones obtained by interpolating those of Ref.~\cite{Athenodorou:2021qvs}, see Figs.~\ref{fig:plot_results}. This is a strong indication that, even when focusing on channels with the same quantum numbers $\mathrm{PC}=-+$ as the topological charge (for example, the $0^{-+}$ or the $2^{-+}$ channels), no systematic error on glueball mass determinations related to topological freezing can be appreciated within our $\sim2-5\%$ level of accuracy.

As a final comment, we observe that our error bars are generally larger than those reported in Ref.~\cite{Athenodorou:2021qvs}, especially for heavier states. This is related to the procedure adopted to extract glueball masses. Indeed, our uncertainties are dominated by systematic effects related to the exponential fit of Eq.~\eqref{eq:correlator_mass_fit}.

In principle, the mass $m$ should be obtained by fitting the large-$t$ asymptotic behaviour of $C_{\mathrm{best}}(t)$ with a single exponential as in Eq.~\eqref{eq:correlator_mass_fit}. In practice, the contamination by larger-mass states at small $t$ and the effects of statistical noise at large $t$ hinder this procedure and produce sizable systematic errors. The choice of the fitting range $[t_{\mathrm{min}},t_{\mathrm{max}}]$ is thus crucial and is determined as follows. We look for a plateau in the effective mass in Eq.~\eqref{eq:effective_mass}, which should signal that the single-exponential asymptotic regime has set in. If a plateau can be identified over an interval $[t_1,t_2]$, we set $t_{\mathrm{min}}=t_1$. The value of $t_{\mathrm{max}}$ is then chosen as the largest $t\le t_2$ which allows to obtain a single-exponential best fit in $[t_{\mathrm{min}},t_{\mathrm{max}}]$ with a reasonable value of the $\chi^2/\mathrm{dof}$, where reasonable means that the corresponding $p$-value is between $5\%$ and $95\%$. If instead a plateau cannot be clearly identified, we estimate $m$ from an envelope of the quasi-plateau of $m_{\mathrm{eff}}$. We stress that such procedure tends to be harder for states whose mass is close to or above the lattice ultra-violet cut-off. In those cases, the plateau is typically very short, the effects of noise immediately apparent and the systematics more prominent. The net result is that the mass of heavier states tends to be determined less precisely.

The results above were obtained after an expensive computation, as the production of our sample of configurations required a budget of 
approximately $\sim 2.3$M core-hours on the cluster where simulations were run.
Unfortunately, our resources did not allow us to improve our statistics further, so as to reach an accuracy comparable to that of Ref.~\cite{Athenodorou:2021qvs} also for heavier states. Nonetheless, we observe a substantial agreement between our results and those in Ref.~\cite{Athenodorou:2021qvs} also for the latter states, confirming the picture that already emerges for lighter states, where 
our accuracy is mostly of the same order of magnitude as the one achieved in Ref.~\cite{Athenodorou:2021qvs}, as can be appreciated from Tab.~\ref{tab:hpt_results}.

\section{Conclusions}\label{sec:conclusions}
In this letter we applied the PTBC algorithm proposed by M.~Hasenbusch to perform the first determinations of glueball masses on the lattice at large-$N$ without any systematic effect related to topological freezing. We did so in the pure $\SU(6)$ gauge theory, at $a \simeq 0.0938$~fm. The masses of the first few low-lying glueball states were computed from a sample of 20k well-decorrelated configurations. We compared our results with those obtained from simulations performed with standard local algorithms, and thus affected by severe topological freezing. 
No systematic effect related to the non-ergodicity of the standard algorithms was observed in the value of glueball masses within our $2-5\%$ level 
of accuracy.

This is a first robust indication\footnote{A preliminary investigation with OBC for $\SU(7)$ was provided in~\cite{Amato:2015ipe}.} that estimates of glueball masses obtained in a fixed topological sector at large-$N$ can be trusted at up to the few percents level. Moreover, this shows that the PTBC algorithm is a perfectly viable solution to the problem of accurately computing glueball masses at large-$N$ without the effects of topological freezing. This algorithm could be easily adopted in more extensive future studies, both to extend our current results to larger values of $N$ and/or to finer lattice spacings.

Several possible future directions can be explored to further clarify the relationship between glueball mass computations and topological properties.\\
An independent way of probing the sensitivity of glueball masses to the choice of a fixed topological sector is to study their dependence on the dimensionless parameter $\theta$, that couples the global topological charge $Q$ to the standard Yang--Mills action. In particular, the quantity $m_2 \equiv \frac{d^2 m_{\mathrm{glueball}}}{d\theta^2}\left\vert_{\theta=0}\right.$ is expected to control the magnitude of systematics effects related to the restriction of the sample of configurations to a fixed topological sector when computing $m_{\mathrm{glueball}}$~\cite{Brower:2003yx}. The computation of $m_2$ has been tackled from $\theta=0$ simulations for the $0^{++}$ state~\cite{DelDebbio:2004xh}, but only compatible-with-zero determinations have been reported for $N\ge4$. This problem could be re-examined from the point of view of the PTBC algorithm in combination with imaginary-$\theta$ simulations, which have been shown to improve the computation of higher-order terms in the $\theta$-expansion~\cite{Bonati:2015sqt,Bonati:2016tvi,Bonanno:2018xtd,Bonanno:2020hht}. Another possible improvement could be achieved by introducing topological operators in the variational basis used in the GEV method, which could help in detecting any possible coupling of glueball states to topological modes. Finally, intriguing insights might come from the application of recent Neural Network techniques~\cite{PhysRevResearch.2.033499}, for instance by training a Neural Network to distinguish correlators computed from samples of configurations with different global topology.

\section*{Acknowledgements}
The authors thank A.~Athenodorou and T.~DeGrand for useful discussions.

C.~B.~acknowledges the support of the Italian Ministry of Education, University and Research under the project PRIN 2017E44HRF, ``Low dimensional quantum systems: theory, experiments and simulations''.

The work of B.~L.~has been supported in part by the STFC Consolidated Grants No. ST/P00055X/1 and No. ST/T000813/1. B.~L.~received funding from the European Research Council (ERC) under the European Union’s Horizon 2020 research and innovation program under Grant Agreement No.~813942. The work of B.~L.~is further supported in part by the Royal Society WolfsonResearch Merit Award No.~WM170010 and by the Leverhulme Trust Research Fellowship No.~RF-2020-4619.

The work of D.~V.~is partly supported by the Simons Foundation under the program ``Targeted Grants to Institutes'' awarded to the Hamilton Mathematics Institute.

Numerical simulations have been performed on the \texttt{MARCONI} machine at CINECA, based on the agreement between INFN and CINECA, under project INF21\_npqcd. Numerical analyses have been performed on the Swansea University \texttt{SUNBIRD} (part of the Supercomputing Wales project) and \texttt{AccelerateAI} A100 GPU system, which are part funded by the European Regional Development Fund (ERDF) via Welsh Government.

\end{document}